\documentclass{PoS}
\usepackage{amsmath,amsfonts,amssymb,amsbsy}
\usepackage{mathrsfs,latexsym}
\usepackage{graphicx}
\usepackage{xcolor}
\usepackage{booktabs}
\usepackage{multirow}
\usepackage{multicol}
\usepackage{subfig}
\usepackage{placeins}
\usepackage{blindtext}
\usepackage{bm}
\usepackage{times}
\usepackage{float}
\usepackage{booktabs}
\usepackage{wrapfig}
\usepackage[utf8]{inputenc} 
\usepackage[T1]{fontenc} 
\usepackage{amsmath,amssymb}
\usepackage{url}
\usepackage{macros_alpha}

\newcommand{\beq}{\begin{equation}}
\newcommand{\eeq}{\end{equation}}
\newcommand{\bea}{\begin{eqnarray}}
\newcommand{\eea}{\end{eqnarray}}
\newcommand{\bean}{\begin{eqnarray*}}
\newcommand{\ean}{\end{eqnarray*}}

\newcommand{\non}{\nonumber}
\newcommand{\mR}[1][]{m_{\mathrm{\sss R}#1}}
\newcommand{\sss}{\scriptscriptstyle}
\newcommand{\Mq}{M_{\rm q}}

\title{Scale setting for QCD with $N_f=3+1$ dynamical quarks}

\ShortTitle{Scale setting for QCD with $N_f=3+1$ dynamical quarks}

\author{\speaker{Roman H{\"o}llwieser}, Francesco Knechtli, Tomasz Korzec\\
Dept. of Physics, University of Wuppertal, Gau{\ss}strasse 20, 42119 Germany}

\abstract{
We present first results of the scale setting for QCD with $N_f=3+1$ dynamical quarks on the lattice. We use a recently proposed massive renormalization scheme with a non-perturbatively determined clover coefficient. To relate the bare coupling of the simulations to a lattice spacing in fm, we measure $t_0^*/a^2$, the flow scale $t_0$ at a  
mass point with $m_\mathrm{up}=m_\mathrm{down}=m_\mathrm{strange}$ and a physical charm quark mass, and assume that $\sqrt{8t_0^\star} = 0.413(5)(2)$fm, as determined in~\cite{Bruno:2017gxd,Bruno:2016plf}. 
We discuss the setup, tuning procedure, simulation parameters and measurement results for ensembles with three different volumes and present a charmonium spectrum.}

\FullConference{37th annual International Symposium on Lattice Field Theory\\
                 16-22 June 2019\\
                 Wuhan, China}

\begin{document}

\maketitle

\section{Introduction}

The omission of a dynamical charm quark from QCD simulations has been shown to have
only little effect on low energy observables~\cite{Knechtli:2017xgy}, but can affect  
quantities with valence charm quarks at a few percent level~\cite{Cali:2019enm}. Moreover, when the strong  
coupling is determined on the lattice in $\Nf=3$ QCD, perturbation theory at the scale of the  
charm quark is necessary to relate it to the phenomenologically relevant $\Nf=5$ result. This can introduce an  
error of up to $1.5\%$ on the $\Lambda$ parameter~\cite{Athenodorou:2018wpk}. A new action with a novel $O(a)$ improvement scheme, specially tailored towards simulations including a charm quark, has been proposed in~\cite{Fritzsch:2018kjg}. We report on first large volume simulations using this action, and in particular concentrate on setting the scale.

The goal is to map out the relation between the bare coupling $g_0$ and  
the lattice spacing $a$ in fm. This relation is, up to lattice artifacts, independent of the quark  
masses. The standard procedure is to determine an experimentally accessible dimensionful quantity at  
the physical mass point in lattice units, and obtain the lattice spacing by using the  
experimental input. This usually requires simulations of whole chiral trajectories at each lattice spacing. We  
propose a method for scale setting that is orders of magnitude cheaper and requires only simulations at  
the flavor $SU(3)$ symmetric point, where the three light quark masses are equal and
\begin{eqnarray}
   \phi_4 &\equiv& 8t_0\left(m_{\text{K}}^2 + \frac{m_\pi^2}{2} \right) = 12t_0m_{\pi,\text{K}}^2=1.11\, , \label{eq:phi4} \\
   \phi_5 &\equiv& \sqrt{8t_0}\left(m_{\text{D}_{\text{s}}} + 2 m_{\text{D}} \right) = \sqrt{72t_0}m_{\text{D,D}_{\text{s}}} = 11.94 \, , \label{eq:phi5} 
\end{eqnarray}
At this mass point $\sqrt{8t_0^\star} = 0.413(5)(2)$fm has been determined in~\cite{Bruno:2017gxd,Bruno:2016plf}. Due to  
decoupling it has the same value in the $3+1$ flavor theory up to a couple per mille, as long as the  
fourth quark's mass is at least as heavy as a charm quark, but this is what is enforced by the second condition Eq.~\ref{eq:phi5}.

Once the relation between $g_0$ and $a$ is mapped out on this particular  
mass point, one can proceed constructing chiral trajectories, {\it e.g.}, along lines where $\phi_4$ and $\phi_5$  
are constant. But already the $SU(3)$ symmetric ensembles are highly useful. They can be  
the starting point for the determination of fundamental parameters of QCD, but also can be  
used directly for charm physics, where the unphysical light quark masses play only a small role.

\section{Simulation setup and scale setting}\label{sec:setup}

Renormalization and improvement conditions are imposed at zero quark mass in
non-perturba\-tive mass-independent schemes. This has many advantages. $Z$-factors
and improvement coefficients depend
only on the bare coupling and renormalization scale, and some perturbative
coefficients, like $b_0$ and $b_1$ in the expansion of the beta-function, are
scheme independent. The drawback however is, that the $O(a)$ improvement pattern
with Wilson fermions can become very complicated. For example, the  
renormalized $O(a)$ improved quark mass is given by~\cite{Fritzsch:2018kjg}
\begin{align}
\mR[,i] &= Z_m(\gtilde^2, a\mu) \bigg[ \mqi + \left( r_m(\gtilde^2) - 1 \right) \frac{\tr[\Mq]}{\Nf} + a \bigg\{
        b_m(g_0^2) \mqi^2 + \bar{b}_m(g_0^2) \tr[\Mq]\, \mqi \nonumber\\
  &\qquad + \left( r_m(g_0^2) d_m(g_0^2) - b_m(g_0^2) \right) \frac{\tr[\Mq^2]}{\Nf} + \left( r_m(g_0^2) \bar{d}_m(g_0^2) - \bar{b}_m(g_0^2) \right) \frac{\left(\tr[\Mq]\right)^2}{\Nf} \bigg\} \bigg] \,, \label{mri0}\\
  \gtilde^2 &= g_0^2 \left( 1 + a\bg(g_0^2) \,\tr[\Mq]/\Nf \right).\non
\end{align}

Most of the improvement coefficients ($b_m$, etc.) are not known beyond  
1-loop of perturbation theory.
This can lead to large uncanceled $O(a)$ effects, if $a\tr(M_q)$ or $a\mqi$ are  
large.
To avoid this complicated improvement pattern, one can give up the  
mass-independence of the scheme.
This allows to absorb all $b$ and $d$ terms into the definition of the,  
now mass dependent,
renormalization factors. A scheme to do so was proposed in~\cite{Fritzsch:2018kjg} and the  
mass dependent clover coefficient $c_{sw}$ in the clover action term 
$S_{\rm SW} = a^5 c_{\mathrm{sw}}(g_0^2,\Mq) \sum_x \bar{\psi}(x) \frac{i}{4} \sigma_{\mu\nu} \hat{F}_{\mu\nu}(x) \psi(x)$
has been determined non-perturbatively. We  
apply this action for the first time
to large volume simulations with a physical charm quark mass. For a first estimate of the bare coupling and quark masses we use the tuning results in~\cite{Fritzsch:2018kjg}, determined on a line of constant physics (LCP). For our first simulation we choose a bare coupling $\beta = 3.24$, light quark masses given by $\kappa_{u,d,s} = 0.134484$ and a charm quark mass by $\kappa_c = 0.12$.
For the algorithmic parameters, we started with the setup of CLS's H400 simulation, cf.~\cite{Bruno:2014jqa}, to which we added the charm quark. The new contribution to the action was not further factorized and the corresponding forces were integrated on the second level of our three level integrator. For our simulations we use openQCD version 1.6\footnote{ M.~L\"uscher, S.~Schaefer, \url{http://luscher.web.cern.ch/luscher/openQCD/}}~\cite{Luscher:2012av} with open boundary conditions in time direction and twisted-mass reweighting, 2nd and 4th order OMF integrators~\cite{OMF}, SAP preconditioning and low-mode-deflation based on local coherence~\cite{Luscher:2003qa,Frommer:2013fsa}. For a full specification of the action with open boundary conditions we also need $c_0=5/3$ for the L\"uscher--Weisz action, boundary improvement coefficients $c_F = c_G = 1.0$ and the clover coefficient from the fit formula~\cite{Fritzsch:2018kjg}
\bean
c_{\rm sw}(g_0^2=6/3.24) &= \frac{1+A g_0^2+B g_0^4}{1+(A-0.196)g_0^2}= 2.18859\,,\quad A = -0.257 \,,\quad B = -0.050\,.
\ean
The u/d quark doublet is simulated with a weight proportional to
   $\text{det}[(D_{oo})^2 ]\frac{\text{det}[\hat D^\dagger \hat D + \mu^2]^2}{\text{det}[\hat D^\dagger \hat D + 2\cdot \mu^2]}$
in terms of the even--odd preconditioned Dirac operator $\hat D$. The strange and charm quarks are simulated with RHMC, and the two rational functions have degrees 12 and 10, respectively with ranges optimized during the tuning process. 
Both the doublet and the rational parts need reweighting and are further factorized according to~\cite{Hasenbusch:2001ne}
\bean
\det[\hat D^2+\mu^2] = \det[\hat D^\dagger \hat D + \mu_0^2] \times \frac{\det[\hat D^\dagger \hat D + \mu_1^2]}{\det[\hat D^\dagger \hat D + \mu_0^2]}
   \times \ldots \times \frac{\det[\hat D^\dagger \hat D+\mu^2]}{\det[\hat D^\dagger \hat D + \mu_{N}^2]},
\ean
such that we have 13 pseudo-fermion fields and 14 actions in total. 

After thermalization on spatially smaller lattices and subsequent doubling of the
spatial dimensions, flow observables and meson masses were computed on a
more-or-less thermalized subset of configurations. It turned out that
the desired tuning point was missed by quite a bit. What makes the tuning process non-trivial is the fact that in $\phi_4$ and $\phi_5$ the mass dependence of $t_0$ and the meson masses go in opposite directions. 
The final tuning point turns out to be 
\bean
\kappa_{u,d,s} &=& 0.13440733,\quad\kappa_c = 0.12784.
\ean

\begin{figure}[h]
   \centering
   \includegraphics[width=0.495\linewidth]{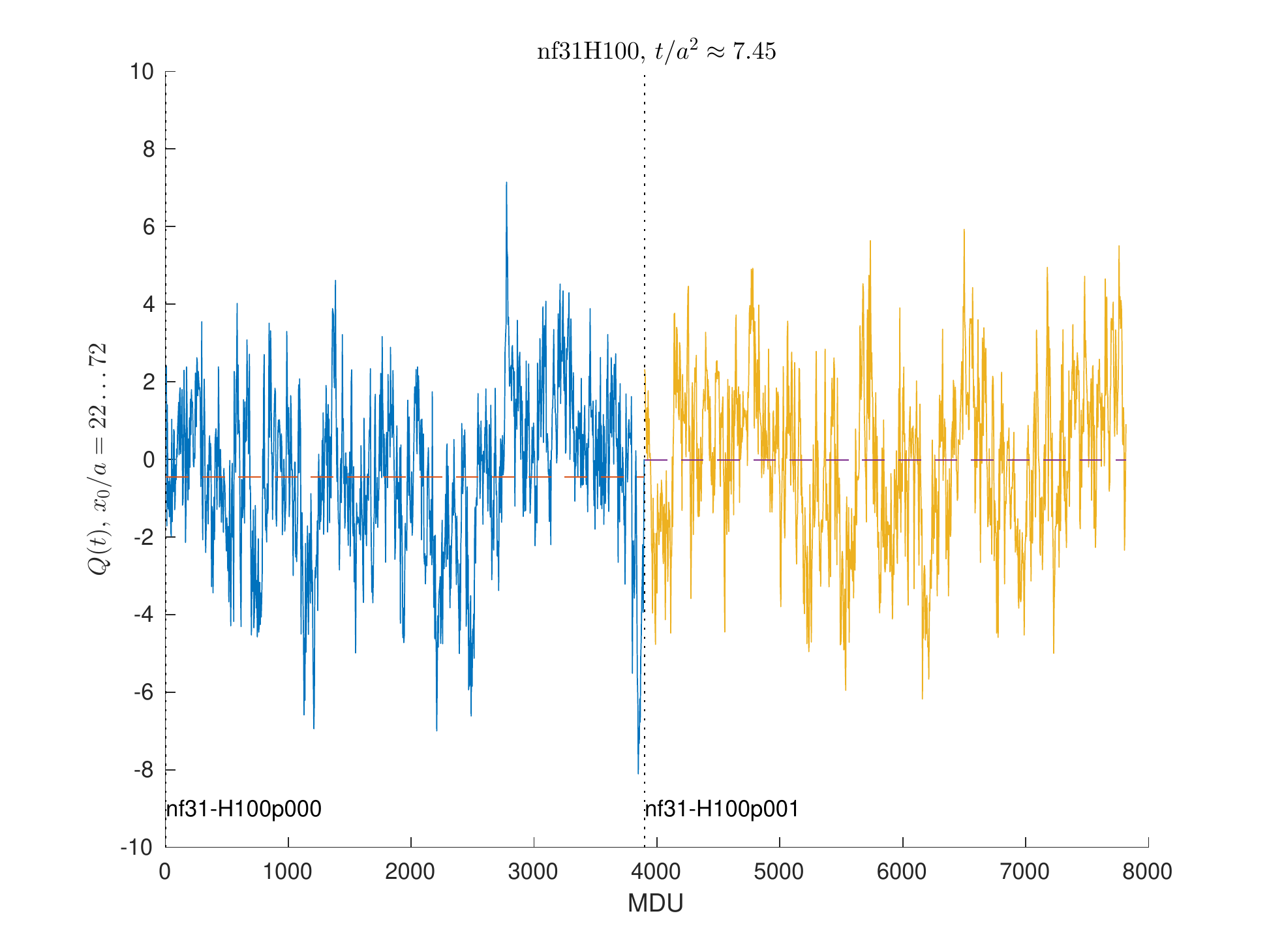} 
   \includegraphics[width=0.495\linewidth]{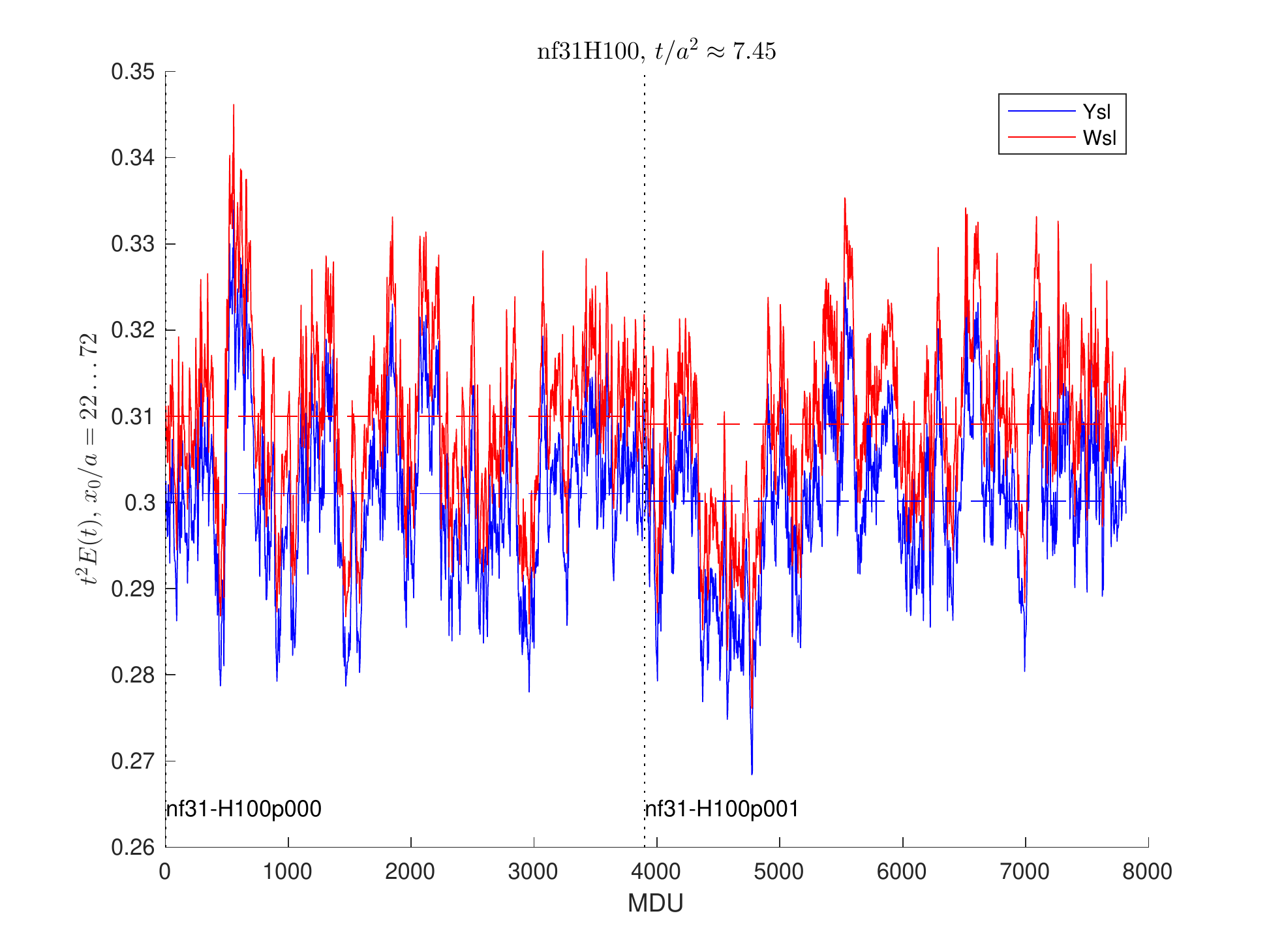}
\caption{Histories of the topological charge $Q(t)$ (left) and of $t^2E(t)$ (right) where $E(t)=\frac{1}{4} G_{\mu\nu}^a(t)G_{\mu\nu}^a(t)$ is the action density of the flowed gauge field, where $t$ corresponds approximately to $t_0$. The two curves show the results for different discretizations of $E(t)$, symmetric (blue) and plaquette (red), which differ by $O(a^2)$ effects. Both replica of ensemble A1 are shown, due to open boundary conditions we average only over time slices $x_0/a=22\ldots72$, as indicated on the ordinate.} 
   \label{fig:histories}
\end{figure}

With these final parameters we produced two high statistics ensembles A1 and A2 with two different lattice sizes given in  Table~\ref{tab:sim} and a short ensemble A0 on a smaller lattice to study finite size effects.  A computation of $t_0^\star/a^2$ on these ensembles, together with the known value of $\sqrt{8t_0^\star} = 0.413(5)(2)$fm~\cite{Bruno:2017gxd,Bruno:2016plf}, yields values for the lattice spacings in fm.  A setup with open boundaries in the temporal direction and periodic boundaries in spatial directions allows us to reach fine lattice spacings~\cite{Luscher:2011kk}, which is crucial for simulations with a dynamical charm quark in the sea. The measurements of the mesonic two-point functions were carried out with the open-source (GPL v2) program ``mesons"~\footnote{T.~Korzec, \url{https://github.com/to-ko/mesons}}, the degenerate pion/kaon and $D$-/$D_s$-meson masses in lattice units as well as our final tuning parameters are also shown in Table~\ref{tab:sim}.  
The results of flow measurements and topological charges are presented in Table~\ref{tab:meas}. The integrated autocorrelation time of $t_0$ is $\tau_{{\rm int}, t_0} \approx 20 \pm 10 \ \text{[4 MDU]}$. Assuming decoupling, {\textit i.e.}, $t_0^\star|_{N_f=3+1}=t_0^\star|_{\Nf=3}+O(1/m_\mathrm{charm}^2)$, our value of $t_0/a^2\approx7.4$ corresponds to a lattice spacing $a \approx 0.054$ fm. The physical size of our $L/a=32$ lattice is $L\approx 1.73$ fm with $m_\pi L = 3.5$, which is a bit small, but finite size effects seem to be under control, as the comparison with $L/a=48$ shows. 
Fig.~\ref{fig:histories} presents example histories of  the action density of the flowed gauge field and the topological charge.

\begin{table}[h]
\centering
\begin{tabular}{cccccccc}
\toprule  
ens. & $\frac{T}{a} \times \frac{L^3}{a^3}$ & $Lm_\pi^\star$ & $N_{traj}$ (MDUs) &$am_{\pi,K}$ & $am_{D,D_s}$ & $\phi_4$ & $\phi_5$ \\
\midrule
A0 & $96 \times 16^3$ & 1.75 & 1400 (3800) & 0.310(6) & 0.614(17) & 10.22(90) & 15.48(43)\\
\midrule
A1 & $96 \times 32^3$ & 3.5 & 3908 (7816) & 0.1137(8) & 0.5247(7) & 1.159(17) & 12.168(40)\\
\midrule
A2 & $128 \times 48^3$ & 5.3 & 3868 (7736) & 0.1107(3) & 0.5228(4) & 1.087(6) & 12.059(20) \\
\bottomrule
\end{tabular}
 \caption{Lattice sizes, statistics and tuning results of the three ensembles with bare coupling $\beta = 3.24$ corresponding to $a=0.054$fm, and quark masses $\kappa_{u,d,s} = 0.13440733$ and $\kappa_c = 0.12784$.}
 \label{tab:sim}
\end{table}

\begin{table}[h]
\centering
\begin{tabular}{cccccccccc}
\toprule  
ens. & $N_{\rm ms}$ & $t_0/a^2$ & $\tau_{int,t_0}$ & $t_c/a^2$ & $\tau_{int,t_c}$ & $w_0^2/a^2$ & $\tau_{int,w_0}$ & $Q^2$ & $\tau_{int,Q^2}$\\
\midrule
A0 & 700 & 8.83(23) & 10(2) & 4.12(9) & 9(4) & - & - & 0.83(11) & 6(1) \\ 
\midrule
A1 & 1954 & 7.43(4) & 16(7) & 3.88(1) & 11(4) & 10.26(13) & 24(12) & 1.08(4) & 5(1) \\ 
\midrule
A2 & 1934 & 7.36(3) & 27(15) & 3.86(1) & 19(12) & 10.13(8) & 30(17) & 6.60(19) & 5(2) \\
\bottomrule
\end{tabular}
 \caption{Flow measurement results and topological charge with integrated autocorrelation times.}
 \label{tab:meas}
\end{table}

\section{Charmonium spectrum and finite size effects}\label{sec:spectrum}

\begin{figure}
   \centering
   \includegraphics[width=.74\linewidth]{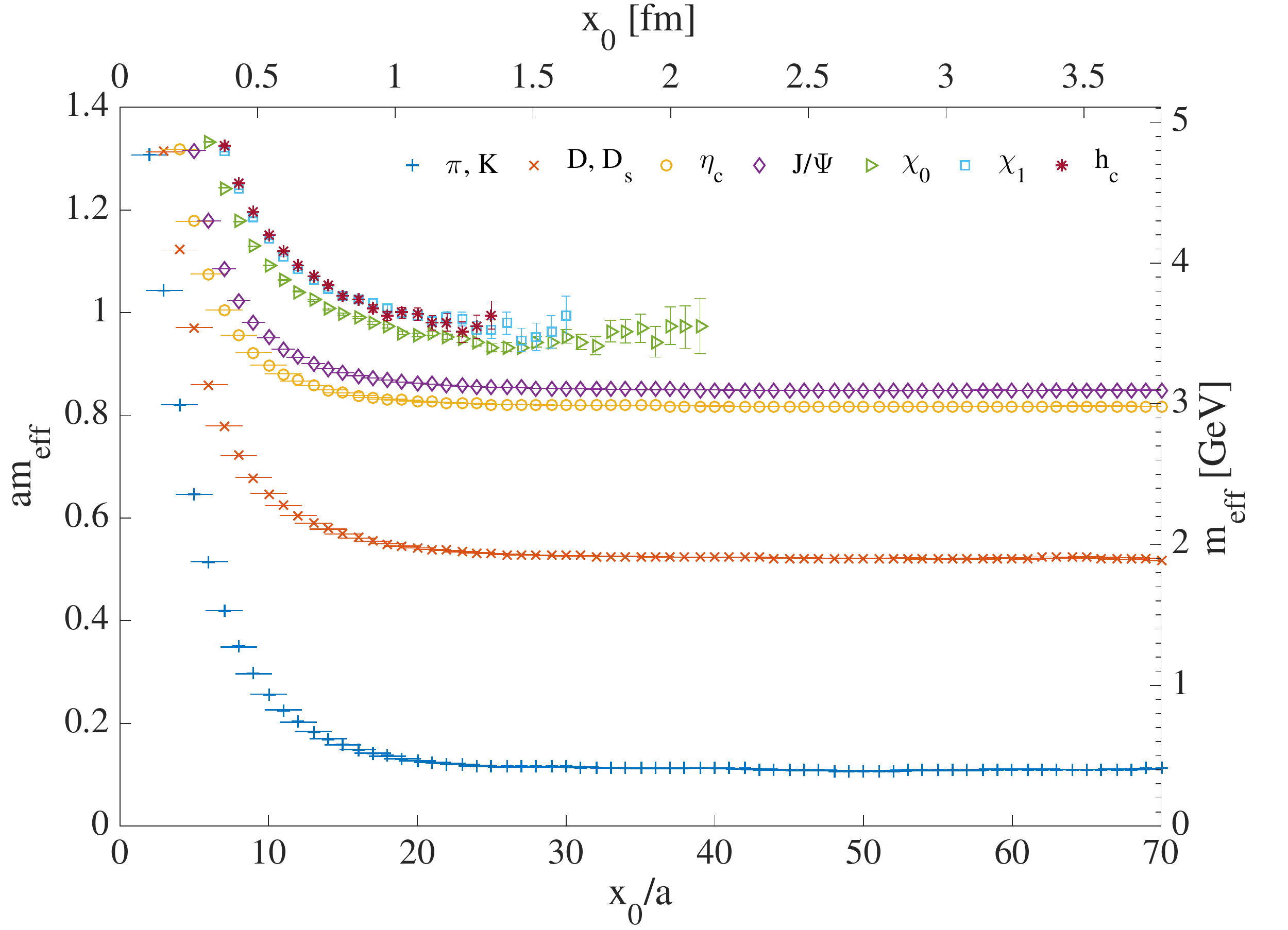}
   \caption{Effective masses of the pion/kaon, $D$- and $D_s$-meson, charmonium states $\eta_c$, $J/\Psi$, $\chi_0$, $\chi_1$ and $h_c$ (from bottom to top).}
   \label{fig:meff}
\end{figure}

In figure~\ref{fig:meff} we show the meson spectrum of our ensemble A2. We get a very clear signal up to the $J/\Psi$ state and can extract reasonable plateau values for higher charmonium states summarized in table~\ref{tab:charm}. We find good agreement for charmonia with PDG data because they contain only charm valence quarks which in our simulations have their physical mass value. Further, the sum of the degenerate light quark masses is at its physical value and since there are no light quarks in the valence sector, the derivatives of the charmonium masses with respect to light quark masses are equal, {\it i.e.} $dm_{x}/dm_{\rm up} = dm_{x}/dm_{\rm down} = dm_{x}/dm_{\rm strange}$. If we want to correct the degenerate light quark masses to their physical values via $m^{\rm phys}_{x} = m_{x} + (\Delta_{\rm up}+\Delta_{\rm down}+\Delta_{\rm strange})\frac{dm_x}{dm_u} + \mathrm{O}(\Delta^2)$, it is clear that the linear term vanishes, because $\phi_4$ is chosen such that $\Delta_{\rm up}=\Delta_{\rm down}=-0.5\Delta_{\rm strange}$ ($m_{u,d,s}=\sum_{i=u,d,s}m_i^{\rm{phys}}/3$) and we only have $\mathrm{O}(\Delta^2)$ corrections.

\begin{table}[b]
   \centering
\begin{tabular}{cccccc}
\toprule  
 & $\eta_c$ & $J/\psi$ & $\chi_{c_0}$ & $\chi_{c_1}$ & $h_c$ \\
\midrule
$am_{eff}$ & 0.8180(2) & 0.8489(2) & 0.9398(86) & 0.9833(72) & 0.9902(81) \\
\midrule
$m_{eff}$ [GeV]  & 2.9890(7) & 3.1019(7) & 3.434(31) & 3.593(26) & 3.618(30) \\
\midrule
PDG [GeV] &  2.9834(5) & 3.096900(6) & 3.4148(3) & 3.51066(7) & 3.52538(11) \\
\bottomrule
\end{tabular}
\caption{Effective masses of charmonium states together with their PDG values.}\label{tab:charm}
\end{table}

\begin{figure}
   \centering
   \includegraphics[width=0.66\linewidth]{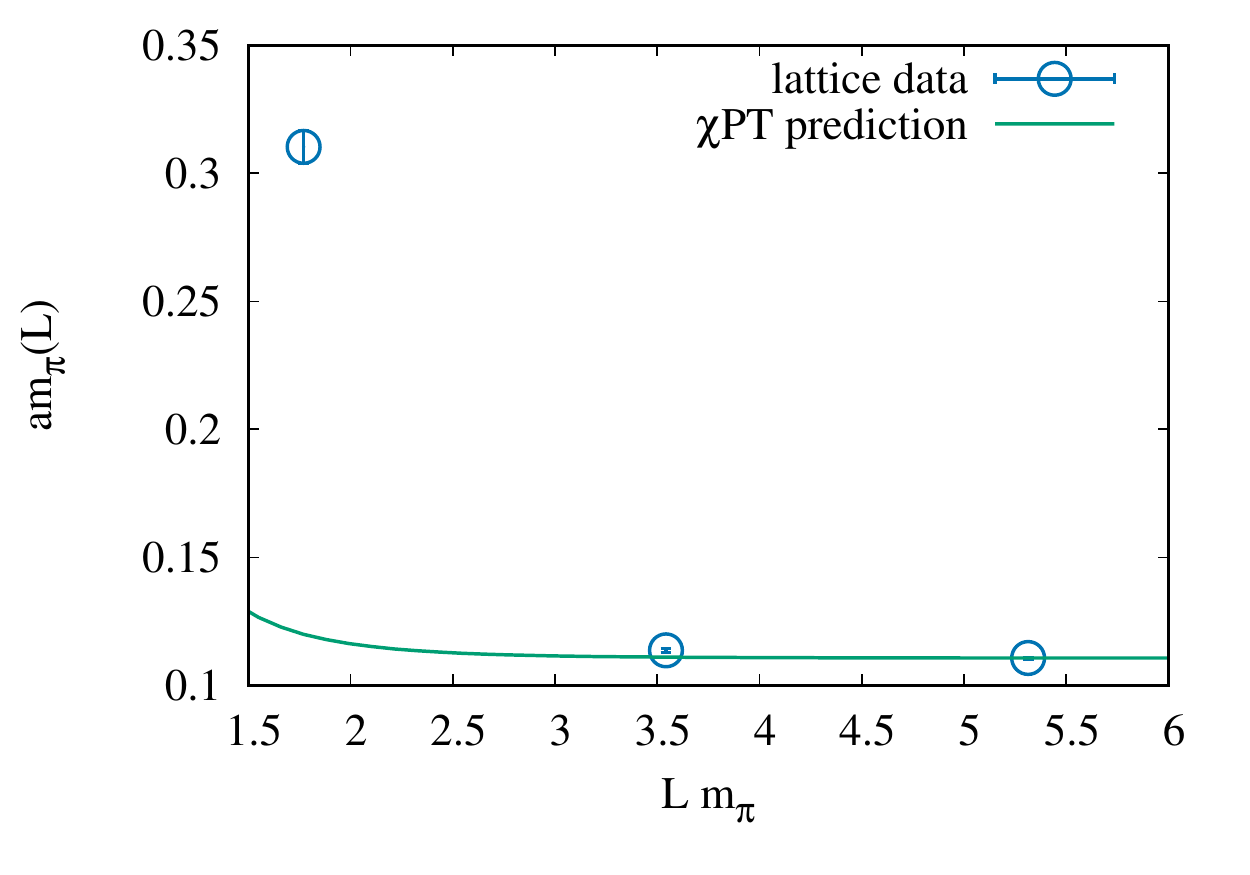}
   \caption{Finite volume scaling effect of $am_\pi$ with $\chi$PT formula in Eq.~\ref{eq:chiptfs} from \cite{Luscher:1985dn,Colangelo:2005gd}} 
   \label{fig:ampi}
\end{figure}

Next, we study finite volume effects of $am_\pi$ following \cite{Luscher:1985dn,Colangelo:2005gd}, who propose an analytic scaling formula from chiral perturbation theory in the p-expansion
\beq
m_\pi(L)=m_\pi\bigg[1+\dfrac{\xi_\pi\tilde g_1(Lm_\pi)}{2\Nf}+\mathcal{O}(\xi_\pi^2)\bigg],\;\tilde g_1(x)=\sum_{n=1}^\infty\dfrac{4m(n)}{\sqrt{n}x}K_1(\sqrt{n}x),\;\xi_\pi=\dfrac{m_\pi^2}{(4\pi F_\pi)^2}.\label{eq:chiptfs}
\eeq
The analytic $\chi$PT prediction together with our lattice data is presented in Fig.~\ref{fig:ampi}, for the pion mass and decay constant at the SU(3) flavor symmetrical point we take the values determined on the finest lattice in~\cite{Bruno:2016plf}. In the range of pion masses and volumes considered the agreement between the one-loop analytical prediction and our lattice data is poor, especially for $Lm_\pi<3.5$.

\section{Conclusions \& Outlook}\label{sec:concl}

We presented the scale setting and tuning of $\Nf=3+1$ QCD using a massive renormalization scheme with a non-perturbatively determined clover coefficient from~\cite{Fritzsch:2018kjg}. We produced two ensembles with lattice sizes $96\times32^3$ and $128\times48^3$ and determine the lattice spacing $a=0.054$ fm. As a first physics result, we measure the masses of the charmonium states $\eta_c$, $J/\psi$, $\chi_{c_0}$, $\chi_{c_1}$ and $h_c$, which we find close to their PDG values. We further plan to study decoupling of the charm quark with light quarks on our ensembles, measure the charmonium sigma terms, disconnected quark loop contributions and the strong coupling $\alpha_S$. To approach the continuum limit we need even larger and finer ensembles. The tuning of ensemble B on a $144\times48^3$ lattice at finer lattice spacing ($a\approx0.043$fm) is finished and production started, and a final ensemble C on a $192\times64^3$ at an even finer lattice spacing may also be simulated.

\section*{Acknowledgements}

We thank Rainer Sommer, Stephan D\"urr and Andrei Alexandru for valuable discussions. We gratefully acknowledge the Gauss Center for Supercomputing (GCS) for providing computer time at the supercomputer JUWELS at the J\"ulich Supercomputing Centre (JSC) under GCS/NIC project ID HWU35. R.H. was supported by the Deutsche Forschungsgemeinschaft in the SFB/TRR55.

\bibliographystyle{gcs}

\begin{thebibliography}{1}

\bibitem{Bruno:2017gxd}
  M.~Bruno, M.~Dalla~Brida, P.~Fritzsch, T.~Korzec, A.~Ramos, S.~Schaefer, H.~Simma, S.~Sint, R.Sommer  [ALPHA Collaboration],
  ``QCD Coupling from a Nonperturbative Determination of the Three-Flavor $\Lambda$ Parameter'',
  Phys.\ Rev.\ Lett.\  {\bf 119} (2017) no.10,  102001

\bibitem{Bruno:2016plf}
  M.~Bruno, T.~Korzec, S.~Schaefer,
  ``Setting the scale for the CLS $2 + 1$ flavor ensembles'',
  Phys.\ Rev.\ D {\bf 95} (2017) no.7,  074504

\bibitem{Knechtli:2017xgy}
F.~Knechtli, T.~Korzec, B.~Leder, G.~Moir [ALPHA collaboration], 
``Power corrections from decoupling of the charm quark",
Phys. Lett. {\bf B774} (2017) 649-655

\bibitem{Cali:2019enm}
S.~Cali, F.~Knechtli, T.~Korzec, ``How much do charm sea quarks affect the charmonium spectrum?",
arXiv:hep-lat/1905.12971 (2019)

\bibitem{Athenodorou:2018wpk}
A.~Athenodorou, J.~Finkenrath, F.~Knechtli, T.~Korzec, B.~Leder, M.~Krsti\'c Marinkovi\'c, R.~Sommer [ALPHA collaboration], ``How perturbative are heavy sea quarks", Nucl. Phys. {\bf B943} (2019) 114612

\bibitem{Fritzsch:2018kjg} 
P. Fritzsch, R. Sommer, F. Stollenwerk, U. Wolff [ALPHA Collaboration],
``Symanzik Improvement with Dynamical Charm: A 3+1 Scheme for Wilson Quarks",
  JHEP {\bf 06} (2018) 025

\bibitem{Bruno:2014jqa}
  M.~Bruno {\it et al.} [ALPHA collaboration],
  ``Simulation of QCD with N$_{f} =$ 2 $+$ 1 flavors of non-perturbatively improved Wilson fermions'',
  JHEP {\bf 1502} (2015) 043

\bibitem{Luscher:2012av}
  M.~L\"uscher, S.~Schaefer,
  ``Lattice QCD with open boundary conditions and twisted-mass reweighting'', 
  Comput.\ Phys.\ Commun.\  {\bf 184} (2013) 519, 
   
\bibitem{OMF}
   I.~P.~Omelyan, I.~M.~Mryglod, R.~Folk,
   ''Symplectic analytically integrable decomposition algorithms: classification, derivation, and application to 
   molecular dynamics, quantum and celestial mechanics simulations'',
   Comp.Phys.Commun. 151 (2003) 272

\bibitem{Luscher:2003qa}
  M.~L\"uscher,
  ``Solution of the Dirac equation in lattice QCD using a domain decomposition method'',
  Comput.\ Phys.\ Commun.\  {\bf 156} (2004) 209 and 
  ``Local coherence and deflation of the low quark modes in lattice QCD'',
  JHEP {\bf 0707} (2007) 081
  
  \bibitem{Frommer:2013fsa}
   A.~Frommer, K.~Kahl, S.~Krieg, B.~Leder, M.~Rottmann, ``Adaptive Aggregation Based Domain Decomposition Multigrid for the Lattice Wilson Dirac Operator", SIAM J. Sci. Comput. {\bf 36} (2014) A1581-A1608
   
   \bibitem{Hasenbusch:2001ne}
  M.~Hasenbusch,
  ``Speeding up the hybrid Monte Carlo algorithm for dynamical fermions'', 
  Phys.\ Lett.\ B {\bf 519} (2001) 177
  
  \bibitem{Luscher:2011kk}
  M.~L\"uscher, S.~Schaefer,
  ``Lattice QCD without topology barriers'',
  JHEP {\bf 1107} (2011) 036

\bibitem{Luscher:1985dn}
M.~L\"uscher, ``Volume Dependence of the Energy Spectrum in Massive Quantum Field Theories. 1. Stable Particle States",
Commun. Math. Phys. {\bf 104} (1986) 177

\bibitem{Colangelo:2005gd}
      G.~Colangelo, S.~D\"urr, C.~Haefeli, 
      ``Finite volume effects for meson masses and decay constants",
      Nucl. Phys. {\bf B721} (2005) 136-174

\end{thebibliography}

\end{document}